\documentclass[twocolumn]{aastex62}

\usepackage{txfonts}
\usepackage{bm}

\newcommand{\beq}{\begin{equation}}
\newcommand{\eeq}{\end{equation}}
\newcommand{\beqn}{\begin{eqnarray}}
\newcommand{\eeqn}{\end{eqnarray}}
\renewcommand{\leq}{\leqslant}
\renewcommand{\geq}{\geqslant}
\newcommand{\eqref}[1]{(\ref{#1})}

\newcommand{\dfrac}[2]{ {\displaystyle\frac{#1}{#2}} }

\newcommand{\pfrac}[2]{ \biggl(\dfrac{#1}{#2}\biggr) }

\newcommand{\au}{{\rm au}}

\defcitealias{OMS+16}{O16}

\begin{document}
\title{ Nonsticky Ice at the Origin of 
the Uniformly Polarized Submillimeter Emission from the HL Tau Disk}

\author{Satoshi Okuzumi}
\affiliation{Department of Earth and Planetary Sciences, 
Tokyo Institute of Technology, Meguro, Tokyo 152-8551, Japan}

\author{Ryo Tazaki}
\affiliation{Astronomical Institute, Tohoku University, Sendai 980-8578, Japan}

\correspondingauthor{Satoshi Okuzumi}
\email{okuzumi@eps.sci.titech.ac.jp}

\shortauthors{Okuzumi et al.}

\begin{abstract}
Recent (sub)millimeter polarimetric observations toward the young star HL Tau 
have successfully detected  polarization emission from its circumstellar disk. 
The polarization pattern observed at 0.87 mm is uniform and parallel to the disk's minor axis,
consistent with the self-scattering of thermal emission by dust particles whose maximum radius is $\approx 100~\micron$.
However, this maximum size is considerably smaller than anticipated from dust evolution models
that assume { a high sticking efficiency for icy particles}.
Here, we propose that the unexpectedly small particle size can be explained if 
CO$_2$ ice covers the particles in the outer region of the HL Tau disk. 
CO$_2$ ice is one of the most major interstellar ices, 
and  laboratory experiments show that it is { poorly sticky}.  
{  Based on dust evolution models accounting for CO$_2$ ice mantles as well as aggregate sintering and post-processing radiative transfer, we simulate the polarimetric observation of HL Tau at 0.87 mm. We find that the models with CO$_2$ ice mantles better match the observation. 
These models also predict that only particles lying between the H$_2$O and CO$_2$ snow lines
can grow to millimeter to centimeter sizes, and that their rapid inward drift results in 
a local dust gap similar to the 10 au gap of the HL Tau disk.}
We also suggest that the millimeter spectral index for the outer part of the HL Tau disk 
is largely controlled by the optical thickness of this region and does not necessarily 
indicate dust growth to millimeter sizes.
\end{abstract}

\section{Introduction}\label{sec:intro}
Dust particles in protoplanetary disks are the ultimate building blocks of planets 
and also the primary opacity sources in the disks. 
Understanding how dust grows in the disks is essential to understanding
how planet formation begins.

{ The disk around the young star HL Tau is an interesting object for studying 
how dust particles in young disks grow and evolve.
The HL Tau disk is well known for its spectacular multiple dust rings revealed by ALMA \citep{ALMA+15}, 
but is also known for its detectable polarized emission at millimeter and submillimeter wavelengths \citep{SLK+14,KTP+17,SYL+17}. 
The morphology of the (sub)millimeter polarization patterns of this system is particularly interesting because it is highly wavelength dependent.
The polarization map at 0.87 mm shows a uniformly polarized pattern in which 
the polarization vectors are parallel to the disk's minor axis \citep{SYL+17}. In contrast, the map at 3.1 mm shows a completely different polarization pattern with azimuthally aligned polarization vectors \citep{KTP+17}. The polarization pattern at 1.3 mm is a mix of the unidirectional and azimuthal patterns \citep{SYL+17}, indicating that the uniformly polarized emission component diminishes as the wavelength increases from  0.87 mm.
There is growing evidence that many other disks also produce uniformly polarized emission 
at (sub)millimeter wavelengths \citep{LLC+18,CHL+18,SMS+18,HYL+18,HCL+18,BGP+18,DPC+19,TMT+19}.
}

{ A likely mechanism for producing such uniformly polarized emission 
is self-scattering of thermal radiation by dust particles (\citealt{KMM+15}; \citealt{YLLS16}, 
see also \citealt{SK01}; \citealt{MS11} for a similar polarization mechanism 
for brown dwarfs and self-luminous planets).
This mechanism well explains the wavelength-dependent nature of 
the uniform polarization for the HL Tau disk because dust particles can only efficiently 
produce polarized scattered light at a wavelength similar to their own size \citep{KMM+15}.
\citet{KMM+16,KTP+17} and \citet{SYL+17} conclude that 
the uniformly polarized emission seen in the 0.87 and 1.3 mm images of the HL Tau disk 
are most likely produced by dust particles of radii $\sim 100~\micron$.  
\citet{KMM+16,KTP+17} also conclude that 
millimeter- to centimeter-sized particles should be much less abundant than 100 $\micron$-sized particles, at least in the region where the polarized emission is observed, because such large particles would produce unpolarized thermal emission with little polarized scattered light at these wavelengths \citep{KMM+16,KTP+17}. 
The uniform (sub)millimeter polarization seen in other disks also gives a similar constraint on the maximum particle size (see the references listed in the previous paragraph). }

{ The above interpretation for the uniform (sub)millimeter polarization 
raises the question why the $100~\micron$-sized particles are so prevalent.
Theoretically, it has long been believed that icy grains in the cold outer part of the disks 
are highly sticky thanks to strong hydrogen bonding between H$_2$O molecules \citep[e.g.,][]{CTH93,DT97,WTS+09,WTO+13}. 
Dust growth models assuming a high sticking efficiency for icy particles 
generally predict that the particles should grow to 1 mm or even larger 
 \citep[e.g.,][]{BDH08,BDB10,OTKW12,DD14,BPR+15}, thus unable to explain 
 the observed (sub)millimeter polarization of the HL Tau disk with the dust self-scattering mechanism. 

To summarize, the uniformly polarized (sub)millimeter emission observed toward the HL Tau disk and many other protoplanetary disks imply that the growth of icy particles 
in the outer regions of the disks may not be as efficient as previously anticipated.
It could merely indicate that the previous dust growth 
models underestimated the stickiness of water ice. 
In fact, some recent experiments \citep{GSK+18,MW19} 
question a high adhesion energy for water ice at temperatures below 150--200 K. 
However, these new experiments are apparently inconsistent with earlier experiments \citep{GB15}  
that confirmed efficient sticking of H$_2$O-ice grains at temperatures down to 100 K.
Alternatively, it is possible that 100-$\micron$-sized aggregates made of $\micron$-sized grains are considerably less sticky than the grains themselves. For instance, experiments show that macroscopic aggregates made of silica grains do not stick but bounce off at moderate collision velocities \citep[e.g.][]{GBZ+10}. In principle, aggregates of water ice grains could also experience bouncing at similar velocities if they are compact \citep{WTS+11}. If this is the case, bouncing could limit the growth of icy  particles more severely than fragmentation \citep[e.g.,][]{ZOG+10,ZODH11}. 

}

{ In this study, we explore another possibility that 
the growth of of $100~\micron$-sized particles in the outer part of the disks is
suppressed by a nosticky solid material---CO$_2$ ice. 
Models of interstellar and circumstellar ices suggest that dust grains in dense and cold environments are covered by ``apolar'' ices of CO$_2$ and CO} \citep[e.g.,][]{BGW15}.
Recent laboratory experiments by \citet{MTJW16a,MTJW16b} show that CO$_2$ ice particles 
are considerably less sticky than H$_2$O particles of the same size, 
presumably because of the absence of hydrogen bonding. 
\citet{MTJW16a} and \citet{PPSB17} point out that the CO$_2$ ice mantles can suppress dust growth 
in the outer part of protoplanetary disks.
{ Here, we incorporate the low stickiness of CO$_2$ ice mantles 
into our previous dust evolution model for the HL Tau disk (\citealt{OMS+16}; henceforth \citetalias{OMS+16}), and demonstrate that this effect can indeed cause 
the uniform submillimeter polarimetric pattern seen in the HL Tau disk.
}

{ The structure of this paper is as follows. 
Section~\ref{sec:model} describes our dust evolution model and radiative transfer approach 
used to synthesize disk polarimetric images.
Section~\ref{sec:results} presents main results from the dust evolution calculations and 
compares them with the submillimeter polarimetric observation of the HL Tau disk.
Section~\ref{sec:discussion} presents some implications for the ring--gap substructure of the HL Tau disk
and also discusses model and parameter dependences. Section~\ref{sec:summary} presents a summary.
}

\section{Model}\label{sec:model}
{ 
To produce synthetic polarimetric images of the HL Tau disk,
we perform global dust evolution simulations and post-processing radiative transfer calculations.  
In this section, we describe the assumptions made in the calculations. 
}

\subsection{ Gas Disk Model}\label{sec:gas}
{ Following \citetalias{OMS+16}, we use simple prescriptions for the gas surface density and temperature profiles of the HL Tau disk.} 
The gas surface density is given by 
\beq
\Sigma_g = \frac{(2-\gamma)M_{\rm disk}}{2\pi r_c^2}\pfrac{r}{r_c}^{-\gamma} 
\exp\left[  - \pfrac{r}{r_c}^{2-\gamma} \right],
\label{eq:Sigmag}
\eeq
where $r$ is the distance from the central star, 
$r_c$ and $M_{\rm disk}$ are the characteristic radius and total mass of the gas disk,
respectively, and $\gamma$ is a dimensionless number characterizing the radial slope of $\Sigma_g$. 
{ Equation~\eqref{eq:Sigmag} is motivated by the similarity solution 
for evolving viscous accretion disks \citep{LP74,HCGD98},
although we do not follow the evolution of the gas disk for simplicity. 
The values of $r_c$ and $M_{\rm disk}$ are taken to be  
$150~\rm au$ and $0.1M_\sun$, respectively.

The actual profile of $\Sigma_g$ in the HL Tau disk may not be smooth. 
\citet{YLG+16} suggest that the distribution of HCO$^+$ in the HL Tau disk has radial gaps, indicating that the profile of $\Sigma_g$ (which is dominated by H$_2$) might also have gaps.  \citet{2019arXiv190408899H} show that the combined effects of non-ideal magnetohydrodynamics and dust-dependent ionization chemistry can produce gas gaps at dust gaps. Such complications are not considered in this study.

The value of $\gamma$ is taken to be either 1 or 0. 
The model with $\gamma = 1$ corresponds to the classical viscous accretion disk model with a radially constant  
viscosity-$\alpha$ parameter \citep{HCGD98}.  The model with $\gamma = 0$ has a flat surface density profile at $r \ll r_c$. 
Such a flat profile can be see in some recent simulations of dust accretion including magnetically driven disk winds \citep{SOM+16}. 
In fact, HL Tau is accompanied by strong jets and winds \citep{KMMJ16}, and the winds may significantly contribute to the evolution of the HL Tau disk \citep{HOFT17}.  
We note, however, that the global picture of wind-driven accretion is still uncertain and likely depends 
on the assumed radial distribution of magnetic field strength \citep{B16}. 
In any case, the gas distribution of the HL Tau disk is essentially unknown, 
and therefore it is important to assume different surface density profiles. 
}

The temperature profile is given by 
\beq
T = 310 \pfrac{r}{1~\rm au}^{-0.57} = 30\pfrac{r}{60~\rm au}^{-0.57} ~\rm K.
\label{eq:T}
\eeq 
{ 
\citetalias{OMS+16} derived this profile based on the assumption that the bright central part 
and rings of the HL Tau disk are optically thick at the wavelength of 0.87 mm.  
Similar midplane temperature profiles were also derived by previous radiative transfer models 
for the HL Tau disk  \citep{MHF99,KLMW15}.  
}

{ The disk is assumed to be vertically isothermal,
and the vertical distribution of the gas is given by a Gaussian with the scale height $H_g = c_s/\Omega$, 
where $c_s$ and $\Omega$ are the sound speed and local Keplerian frequency, respectively.
The Keplerian frequency depends on the stellar mass, which we take} to be $1.7M_\sun$ following recent mass estimates \citep{PDM+16,YTC+17}. 

\begin{figure}[t]
\centering
\resizebox{\hsize}{!}{\includegraphics{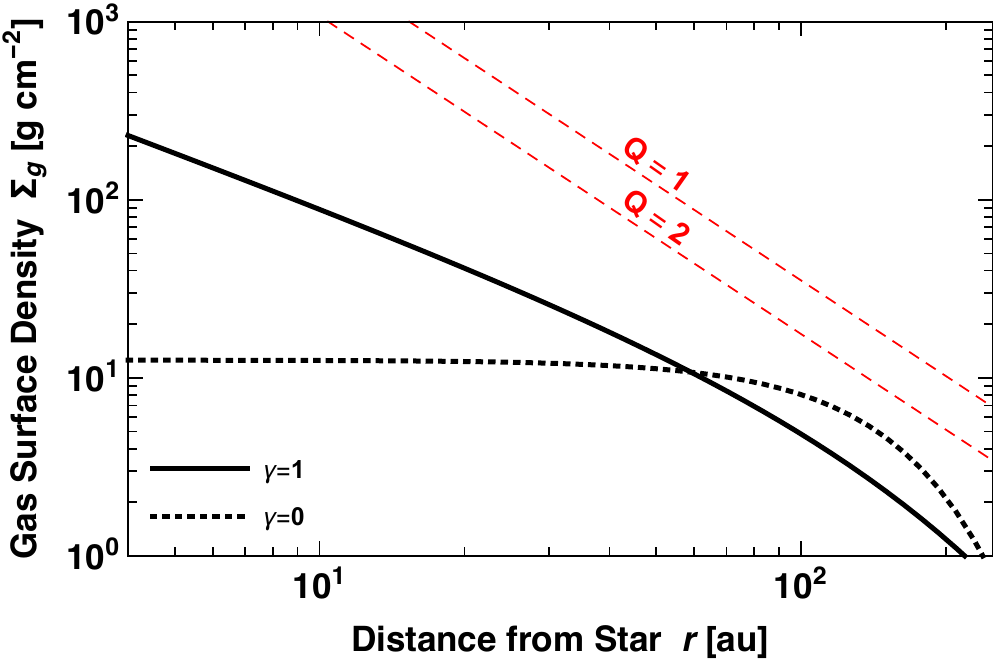}}
\caption{Gas surface density profiles for $\gamma =$ 1 and 0 (solid and dotted curves, respectively). 
The dashed lines mark $Q \equiv c_s\Omega/(\pi G \Sigma_g) =$ 1 and 2. 
Both disk models satisfy $Q \ga 1$ and therefore gravitationally stable. }
\label{fig:Sigmag}
\end{figure}
{ 
Figure~\ref{fig:Sigmag} shows the radial distribution of $\Sigma_g$ for $\gamma = 1$ and 0.
Both disk models are gravitationally stable because the Toomre parameter $Q \equiv c_s\Omega/(\pi G \Sigma_g)$ is greater than unity at all $r$. 
}

\subsection{Dust Model}\label{sec:dust}
We use a modified version of the dust evolution model \citetalias{OMS+16}, 
which was originally developed to explain the multiple dust ring structure of the HL Tau disk.  
{ 
The model tracks the growth, fragmentation, and radial inward drift 
of dust aggregates at different distances from the central star. The single-size approach \citep{SOI16} is adopted, in which aggregates at each radial distance are characterized by a single radius $a_*$.
Effectively, this approach tracks the evolution of the largest, mass-dominating aggregates,  
and the characteristic size $a_*$ corresponds to the size of the largest aggregates.
Aggregates smaller than the largest are neglected in the calculations of dust evolution, but are taken into account in radiative transfer calculations (see Section~\ref{sec:radiative}).} 
Each aggregate is composed of icy dust grains whose composition changes 
with temperature and hence with radial distance. 
The grain composition controls the stickiness of aggregates, thus central to this study. 
Aggregate sintering (\citealt{S11b}; \citetalias{OMS+16}) is also included to reproduce multiple dust rings as seen in the high-resolution image of the HL Tau disk \citep{ALMA+15}. 
 
{ 
In the following subsections, we describe the main aspects of our dust model. 
We refer interested readers to Sections 3 and 4 of \citetalias{OMS+16}
for a more detailed description of the model.
}

\subsubsection{Grain and Aggregate Structure}\label{sec:grain}
Dust particles are assumed to be { initially in the form of micron-sized spherical grains}.
Each grain consists of a silicate core and ice mantles. 
The ice mantles are mainly composed of H$_2$O, CO$_2$, and CO
{ with a molar ratio of 1:0.1:0.1}.
We also consider less abundant CH$_4$ and C$_2$H$_6$ ice 
but only in modeling aggregate sintering (see below).
The initial solid-to-gas mass ratio of the disk is set to 0.01, and the positions of the snow lines for all ices are
computed by comparing their equilibrium vapor pressures and partial pressures in the initial state. { The equilibrium vapor pressures depend on the sublimation energies, which are typically uncertain to $10\%$. Following \citetalias{OMS+16}, we take the sublimation energies of H$_2$O and C$_2$H$_6$ to be $10\%$ lower than the fiducial values to tune the locations of their snow lines within the uncertainty of the sublimation energies (see Section 3.2 of \citetalias{OMS+16} for details).}    
We do not track the evolution of the snow lines for simplicity.

The core--mantle grains stick together to form aggregates. 
{ To distinguish between initial core--mantle grains and their aggregates, we refer to the former as monomers as well as grains. }
The internal density $\rho_{\rm int}$ of the aggregates is taken to be $0.64~\rm g~cm^{-3}$ assuming 
a mean material density of $1.28~\rm g~cm^{-3}$ and an aggregate porosity of 50\%. 
{ In protoplanetary disks, aggregates could have much lower porosities, 
in particular when they grow through mutual collisions without fragmentation \citep{SWT08,OTKW12,KTOW13b}. Some issues with our assumption on the aggregate porosity are discussed in Section~\ref{sec:porous}.

}

{ 
The stickiness of grains and their aggregates depends on 
the materials coating the grain surfaces  \citep{DT97}.
The structure of the ice mantles is therefore of particular importance in our model.
According to the experiments by \citet{MTJW16b}, pure CO$_2$ ice is poorly sticky, 
whereas a homogeneous mixture of H$_2$O and CO$_2$ ices with a ratio of 1:0.1 
would be almost as sticky as pure H$_2$O ice (see their Figure 6).  
The original model of \citetalias{OMS+16} did not account for the low stickiness of 
CO$_2$ ice, and therefore effectively assumed a homogeneous H$_2$O--CO$_2$ ice mantle. 
In this study, we also consider a two-layer mantle model in which the silicate cores are covered with 
an H$_2$O-dominated lower mantle and a CO$_2$-dominated upper mantle, 
similar to models for interstellar ice grains \citep[see, e.g., Figure 10 of][]{BGW15}.
We refer to the former and latter models as the models with and without CO$_2$ ice mantles, respectively.  
The H$_2$O and CO$_2$ ice mantles are assumed to sublimate instantaneously 
at the corresponding snow lines. 
}

\subsubsection{Fragmentation Threshold}
The grain aggregates are assumed to fragment rather than stick if they collide at velocities 
greater than a threshold $v_{\rm frag}$.
{ Bouncing collisions, often observed in laboratory collision experiments 
for rocky aggregates \citep[e.g.,][]{GBZ+10}, are not considered in this study 
to focus on the role of CO$_2$ mantles. 
Our future modeling will take in account bouncing collisions.
}

{ 
The value of $v_{\rm frag}$ is primarily determined by 
the material that covers the grains. 
For aggregates of H$_2$O-mantled grains, we follow \citetalias{OMS+16} in adopting
the following relation between  $v_{\rm frag}$  and $a_{\rm mon}$, 
\beq
v_{\rm frag} = 50\pfrac{a_{\rm mon}}{0.1~\micron}^{-5/6} {\rm m~s^{-1}} 
= 5.2\pfrac{a_{\rm mon}}{1.5~\micron}^{-5/6} ~{\rm m~s^{-1}}.
\label{eq:vfrag_ice}
\eeq
This scaling is based on collision simulations for equal-sized H$_2$O ice aggregates \citep{WTS+09}.  
Equation~\eqref{eq:vfrag_ice} is consistent with the experiments by \citet{GB15}, 
which suggest $v_{\rm frag} \sim 10~{\rm m~s^{-1}}$ for aggregates of $1.5~\micron$-sized H$_2$O-ice grains at 
$T \approx 100$--200 K (see Figure 9 and 10 of \citealt{GB15}; note that there is a factor of $\sim$2 uncertainty in the experimental data shown in these figures).  
For aggregates of bare silicates, 
collision simulations by \citet{WTS+09,WTO+13} show that the sticking threshold is about 10 times 
higher than that for H$_2$-ice aggregates, so we adopt
\beq
v_{\rm frag} = 5\pfrac{a_{\rm mon}}{0.1~\micron}^{-5/6} {\rm m~s^{-1}} 
= 0.52\pfrac{a_{\rm mon}}{1.5~\micron}^{-5/6} ~{\rm m~s^{-1}}.
\label{eq:vfrag_sil}
\eeq
Experiments by \citet{PBH00a} show a capture velocity of $\sim 1~{\rm m~s^{-1}}$ 
for silica grains of $0.6~\micron$ radius ($1.2~\micron$ diameter), 
consistent with Equation~\eqref{eq:vfrag_sil} ($v_{\rm frag} \approx 1.1~{\rm m~s^{-1}}$
for $a_{\rm mon} = 0.6~\micron$). 
For aggregates of CO$_2$-mantled grains,  
experiments by \citet{MTJW16a,MTJW16b} suggest that 
the fragmentation threshold is close to that for silicate grain aggregates, 
and therefore we use Equation~\eqref{eq:vfrag_sil}.
}
 
{ Our model also accounts for a decrease in $v_{\rm frag}$ due to sintering \citep{S99,SU17}. 
We assume that volatiles included in monomer grains are able to desorb from or diffuse over the grain surface when the temperature is close to their sublimation points. 
Sintering refers to the phenomenon in which the mobile volatile molecules recondense around the contact points of the monomers and thereby fuse them together (see, e.g., \citealt{P03,B07}).
This phenomenon makes the aggregates harder but less liable to stick upon high-speed collisions \citep{S99,SU17}. 
Sintering only requires a small amount of materials  \citep[see Figure 2 of][]{SU17} 
and can in principle occur near the snow lines of various volatiles  \citep{S11b}. 
As shown by \citetalias{OMS+16}, the local reduction of the fragmentation threshold in the sintering zones can lead to the formation of the multiple dust rings observed in the HL Tau disk. 
Inclusion of this effect thus allows us to model the submillimeter polarization of this disk consistently with its dust ring structure.
We emphasize, however, that the goal of this study is to explain the submillimeter polarization pattern, not to reproduce the multiple ring structure in full detail. 
We also note that sintering may not be the common origin of the dust rings observed in many protoplanetary disks \citep{LPH18,HAD18,VDD19}.
In any case, our CO$_2$ mantle model for submillimeter polarization is in principle 
compatible with other ring formation mechanisms, such as planet--disk interaction \citep[e.g.,][]{DZW15,DPL+15}, gas--dust instabilities \citep[e.g.,][]{TI14}, and disk dynamics \citep[e.g.,][]{FRD+15}. 

%
Our treatment of sintering is similar to that by \citetalias{OMS+16}, but 
some simplifications are applied given its relatively minor importance in this study. 
We consider sintering by H$_2$O, CO$_2$, C$_2$H$_6$, CH$_4$, and CO. 
Unlike \citetalias{OMS+16}, we neglect NH$_3$ and H$_2$S
because their sintering zones are close to the snow line of of CO$_2$ (in particular, the NH$_3$ sintering zone overlaps with the CO$_2$ snow line). 
Neglecting these species allows us to highlight the effects of CO$_2$ sublimation 
on dust growth near the CO$_2$ snow line.  
The sintering zones are defined by the locations where the timescale 
of sintering is shorter than the mean collision interval. 
In these zones,  
we use the analytic prescription presented in Section 4.4 of \citetalias{OMS+16} to
reduce the value of $v_{\rm frag}$ by up to 60\% \citep{SU17}. 
Note that sintering is assumed to further reduce 
the sticking efficiency of CO$_2$-mantled grain aggregates\footnote{ 
This assumption is based on the finding by \citet{SU17} that the sticking threshold for sintered aggregates is determined by the strength of {\it non-sintered} contacts, and hence by the material that dominates the monomer surface (CO$_2$ ice in this context). When two sintered aggregates collide, they temporary stick by forming new, non-sintered contact points. However, the aggregates immediately separate if the tensile forces acting between the aggregates are large enough to break the non-sintered contacts. The role of sintering here is to hinder the monomers from dissipating kinetic energy through rolling friction. 
}. 
For simplicity, the mean collision interval is approximated by $100\Omega^{-1}$
\citep{TL05,BDH08}. 
This approximation hardly affects the locations of the sintering zones because 
the sintering timescale is a very strong function of $T$ (see Figure 4 of \citetalias{OMS+16}). 

}

{ The fragmentation thresholds as well as the widths of the sintering zones depend 
on the uncertain monomer size $a_{\rm mon}$.  
Our sintering model only requires $a_{\rm mon} \la 4~\micron$ 
for the sintering zones to have a non-zero width \citepalias{OMS+16}. 
Near-infrared polarimetry of HL Tau suggests that the maximum grain sizes in the surrounding envelope  
is more or less $1~\micron$ \citep{LFT+04,MOPI08}, so
it may be reasonable to assume $a_{\rm mon}\sim 1~\micron$.
In this study, we adopt $a_{\rm mon} = 1.5~\micron$ for simulations, 
but also present an analytic estimate of how the simulation results depend on $v_{\rm frag}$.
}

\begin{figure}[t]
\centering
\resizebox{\hsize}{!}{\includegraphics{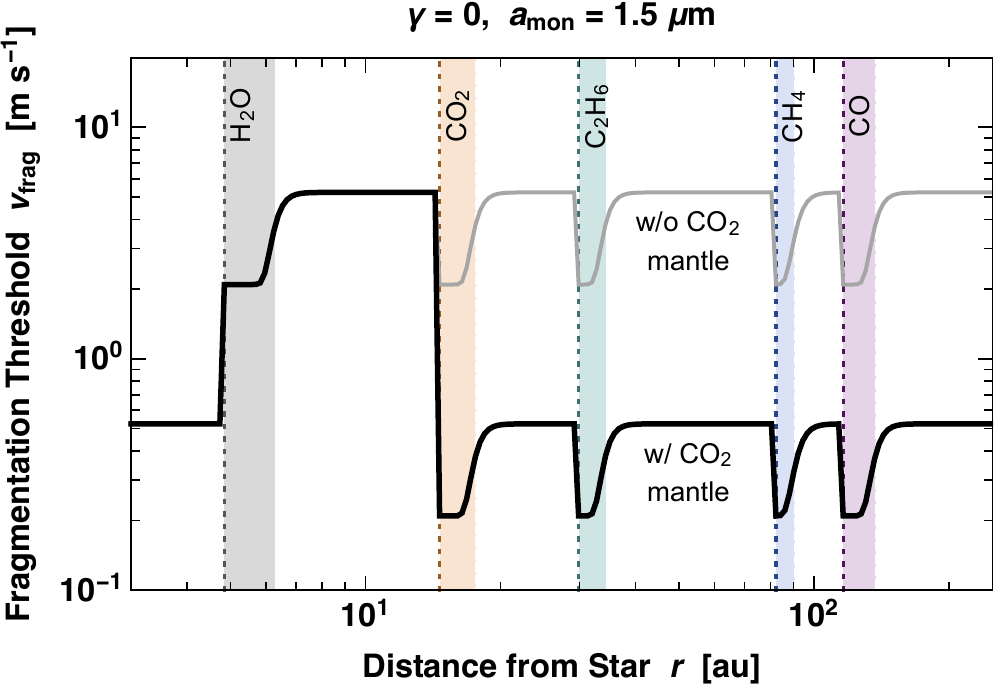}}
\caption{Fragmentation threshold $v_{\rm frag}$ as a function of orbital radius $r$
for models with and without CO$_2$ ice mantles (black and gray curves, respectively).
The vertical dotted lines indicate the locations of the snow lines of five volatile species,
with the shaded regions marking the sintering zones \citepalias{OMS+16}. 
The plot assumes $\gamma = 0$ and $a_{\rm mon} = 1.5~\micron$,
although the results are very insensitive to $\gamma$.
}
\label{fig:vfrag}
\end{figure}

In Figure~\ref{fig:vfrag}, we plot the fragmentation threshold $v_{\rm frag}$ { for $a_{\rm mon} = 1.5~\micron$} 
as a function of $r$ 
for two models with and without CO$_2$ ice mantles.
{ 
In the former model, the fragmentation threshold outside 
the CO$_2$ snow line falls in the range 0.2--$0.5~\rm m~s^{-1}$.
Figure~\ref{fig:vfrag} also shows the locations of the snow lines and sintering zones. 
This locations and widths of the sintering zones are almost the same as 
those in our previous model La0-tuned (see Figure 18 of \citetalias{OMS+16}), 
which adopted the same sublimation energies and a similar monomer size.
Because we neglect NH$_3$ and H$_2$S, our sintering zone at $r \sim 15~\au$
is narrower than in the previous model by a factor of $\sim 2$--3.
}

\subsubsection{Dust Evolution}\label{sec:devol}
{ 
We calculate the time evolution of $a_*$ and the dust surface density $\Sigma_d$ 
as a function of $r$ by solving a set of differential equations with 
advection and collision terms (Equations~(7) and (8) of \citetalias{OMS+16}).  
The radial advection is due to the drift motion of dust in the sub-Keplerian rotating 
gas disk \citep{W72,AHN76,W77a}. 
Turbulent diffusion of dust in the radial direction 
is neglected assuming that turbulence in the HL Tau disk is weak.
This assumption may be justified by the observational evidence 
for a significant level of dust settling in this disk \citep{PDM+16}. 
The computational domain is taken to be $1~{\rm au} \leq r \leq 1000~\rm au$. 
}

{ The motion of an aggregate in the gas disk is characterized by its 
dimensionless Stokes number ${\rm St}$ defined as the product of the aggregate's 
stopping time and $\Omega$.
We use the expression \citep[e.g.,][]{BDB10}
\beq
{\rm St} = \frac{\pi \rho_{\rm int} a_*}{2 \Sigma_g}
\label{eq:St}
\eeq
that  applies to aggregates smaller than the gas mean free path. 
In our simulations, all aggregates fulfill this criterion. 

The collision velocity $\Delta v$ of aggregates is induced by their Brownian motion, 
radial and azimuthal drift \citep{W72,AHN76,W77a}, and gas turbulence. 
We assume that collisions with similar-sized aggregates determine the evolution of 
the largest aggregates,  and take the ratio of the Stokes numbers of two colliding 
aggregates to be 0.5 \citep{SOI16}. 
Analytic expressions by \citet{OC07} are used to calculate
the turbulence-induced collision velocity $\Delta v_t$.
} 

The dust scale height $H_d$ is determined by the balance between 
dust settling and turbulent diffusion \citep{DMS95,YL07}.
{ 
As long as ${\rm St} \ll 1$, $H_d$ can be written as
\beq
H_d = \left( 1+ \frac{{\rm St}}{\alpha_{Dz}} \right)^{-1/2}H_g,
\label{eq:Hd}
\eeq
where $\alpha_{Dz}$ is the vertical diffusion coefficient for dust 
normalized by $c_s^2 / \Omega$.  

The strength of turbulence is parametrized by a dimensionless number $\alpha_{\rm turb}$,
which is defined so that the velocity dispersion of the turbulent gas is given by  
$v_{g, \rm turb} = \sqrt{\alpha_{\rm turb}}c_s$.
The vertical diffusion coefficient for dust is assumed to be  
$\alpha_{Dz} = 0.3 \alpha_{\rm turb}$} 
based on the results of MHD simulations for disk turbulence \citep{OH11}.
We assume a low level of turbulence with $\alpha_{\rm turb} = 3\times 10^{-4}$
to allow dust settling (see Section~\ref{sec:radial}).
{ In our models, the turbulence parameter $\alpha_{\rm turb}$ only affects the collision velocity and vertical diffusion of dust; 
the gas surface density and temperature are given independently of $\alpha_{\rm turb}$
by Equations~\eqref{eq:Sigmag} and \eqref{eq:T}.} 

\subsection{Radiative Transfer}\label{sec:radiative}
The simulation results are converted into polarimetric images using 
the 3D Monte Carlo radiative transfer code \textsc{radmc-3d} \citep{DJP+12}.
{ The dust opacity is calculated using the method detailed in Section 4.5 of \citetalias{OMS+16}. 
In short, we assume that aggregates at each $r$ obeys a power-law size distribution 
whose total surface density and maximum cut-off size are given by $\Sigma_d(r)$ and $a_*(r)$, respectively.} The optical properties of the aggregates are computed 
using the Mie theory combined with the Maxwell--Garnett mixing rule \citep{BH83}.
We use the same effective refractive index for the ice--dust mixture as in \citetalias{OMS+16}.
{ For non-fractal aggregates considered in this study, the absorption and scattering opacities obtained from the effective medium approach agree with those from the more rigorous T-matrix method to an accuracy of  $\la 40\%$ unless the aggregates are much larger than the wavelength
\citep[see Section 4.1.2 of][]{TT18}.} 
{ The inclination angle of the HL Tau disk is assumed to be $46^\circ.7$ \citep{ALMA+15}.  
{Each synthetic polarimetric image is produced using $10^9$ photon packets.}
}

{ We calculate the thermal emission and scattering by dust particles at the wavelength of 0.87 mm 
under the prescribed temperature profile given by Equation~\eqref{eq:T}.  
Because the actual disk temperature profile should depend on the dust distribution in the disk,
our current modeling is not self-consistent. 
Our calculations do not treat radiative transfer of starlight at visible wavelengths, 
and therefore cannot be used to derive the disk temperature profile.
In the particular case of the HL Tau system, modeling the temperature structure with radiative transfer calculations is difficult, if not impossible, because the reflected light from the surrounding envelope can contribute to disk heating \citep{MHF99,KNO02}. 
}

\section{Results}\label{sec:results}

We carried out four simulation runs with and without CO$_2$ ice mantles,
and with $\gamma = 1$ and 0. 
In all simulations, the radial profiles of aggregate size and surface density
relax into a quasi-steady state in which the mass accretion rate of  
inward drifting aggregates stays approximately constant in space (e.g., see Figure 8 of \citetalias{OMS+16}).
This quasi-steady state lasts until the outer part of the disk becomes depleted of dust. 
We find that the quasi-steady state is sustained over $t\sim$1--3 Myr 
and over $t \sim 0.2$--0.4 Myr in the models with and without CO$_2$ ice mantles, respectively,
where $t$ is the time after the start of dust evolution. 
The CO$_2$ mantles slow down dust depletion because they induce collisional fragmentation: smaller aggregates drift more slowly \citep{BDB09}.
In the following, we select the snapshots at $t =0.5~\rm Myr$ and $1.8~\rm Myr$
as the representative results for the models with and without CO$_2$ ice mantles, respectively.

\subsection{Radial Profiles of $a_*$, $\Sigma_d$, and $H_d$}\label{sec:radial}
\begin{figure*}[t]
\centering
\resizebox{8cm}{!}{\includegraphics{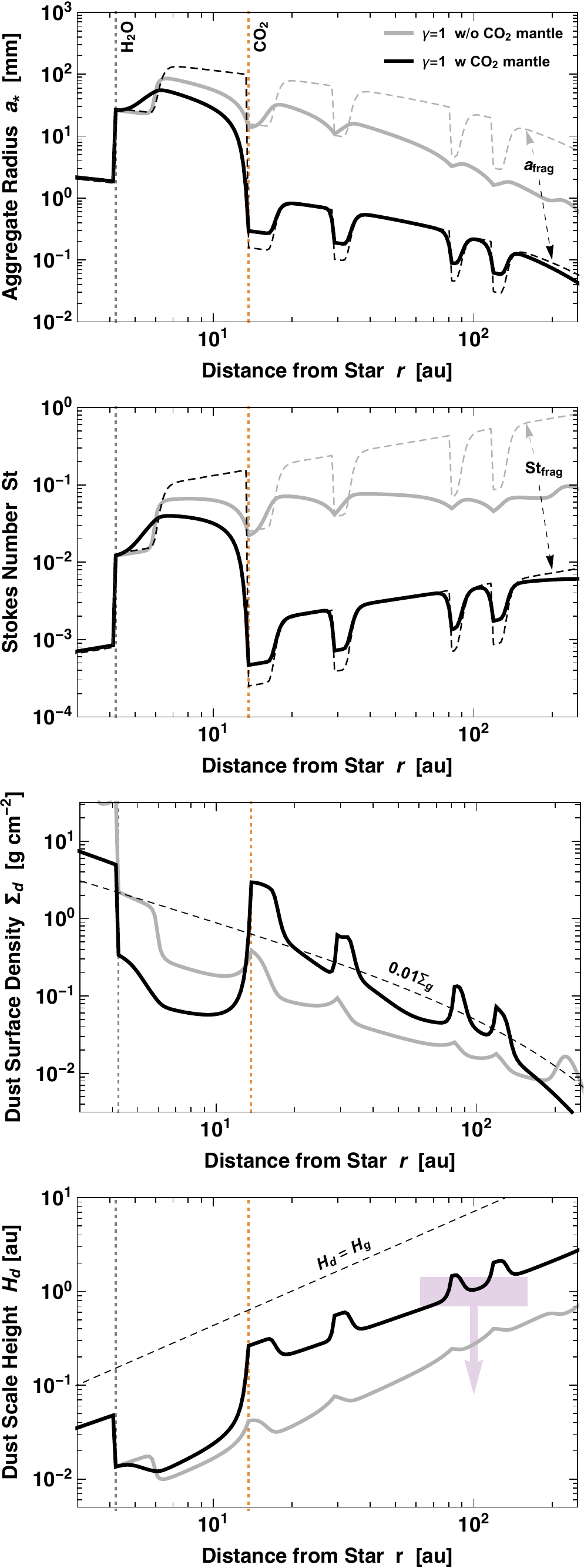}}
\hspace{5mm}
\resizebox{8cm}{!}{\includegraphics{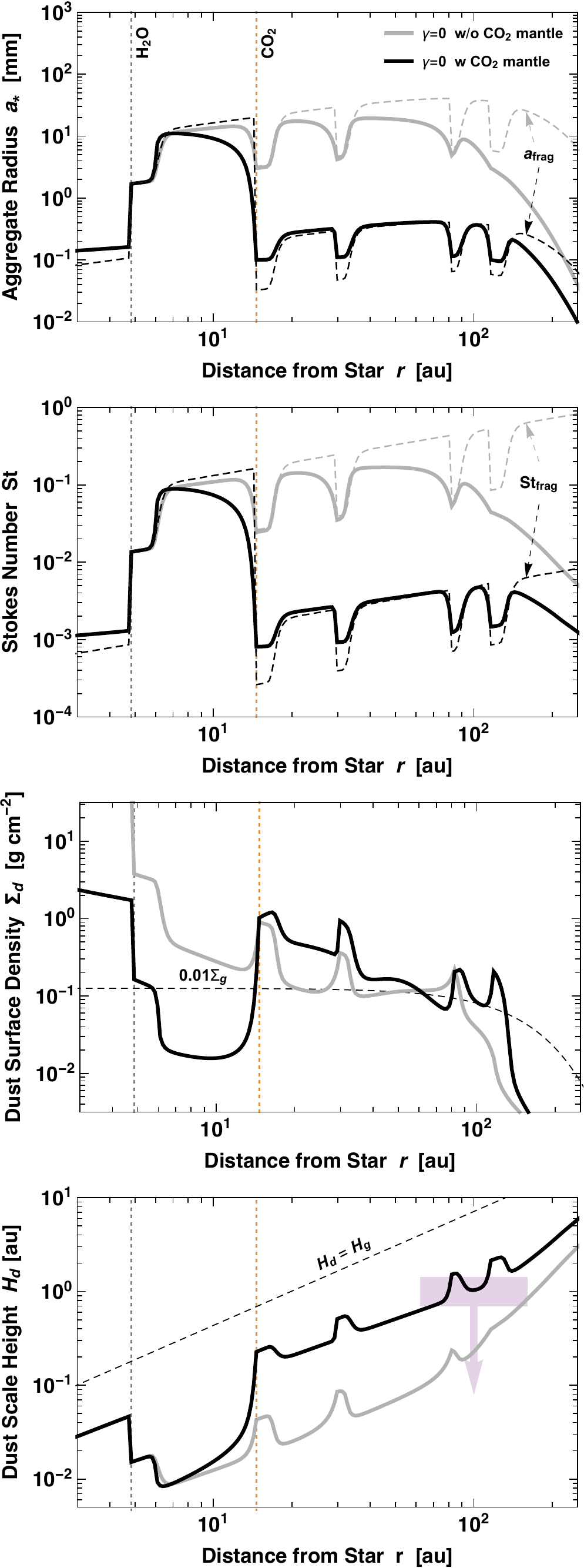}}
\caption{Snapshots of the characteristic size $a_*$ (top row), 
{ Stokes number ${\rm St}$ (second row)}, 
surface density $\Sigma_d$ (third row), and scale height $H_d$ (bottom row)
of dust aggregates as a function of radial distance $r$
for four models with $\gamma = 1$ and 0 (left and right columns, respectively), 
and with and without a CO$_2$ mantle (thick and thin lines, respectively).
The gray and orange vertical lines mark the locations of the H$_2$O and CO$_2$ snow lines.
{ The dashed lines in the panels for $a_*$ and St indicate
the growth limit set by fragmentation (Equations~\eqref{eq:afrag} and \eqref{eq:Stfrag}, respectively). 
The dashed lines in the panels for $\Sigma_d$ and $H_d$ 
mark $\Sigma_d = 0.01 \Sigma_g$ and $H_d = H_g$, respectively.
The purple bars in the bottom panels mark
the upper limit of $H_d$ at $r \approx 100~\rm au$ inferred from 
the geometric thickness of the dust rings in the HL Tau disk \citep{PDM+16}.
}
}
\label{fig:aSigma}
\end{figure*}
The top row of Figure~\ref{fig:aSigma} shows the radial distribution 
of the aggregate size $a_*$ in the quasi-steady state from all simulation runs. 
{ 
We find that the models with CO$_2$ ice mantles successfully produce
 $a_* \sim 100~\micron$ outside the CO$_2$ snow line.
The aggregate size is particularly close to $100~\micron$ in the sintering zones (see Figure~\ref{fig:vfrag})
owing to the combined effects of CO$_2$ mantles and sintering.
}
Without CO$_2$ ice mantles, the aggregates grow beyond 1 mm 
except interior to the H$_2$O snow line and at the outer edge of the disk.

{ 
The results presented here can be understood with simple analytic arguments 
(\citealt{BDB09,BKE12}; \citetalias{OMS+16}). 
In the simulations with CO$_2$ ice mantles, the maximum aggregate size is predominantly 
determined by collisional fragmentation. 
For intermediate-sized aggregates, the collision velocity is mainly induced by 
turbulence, and can be approximately written as $\Delta v \approx \sqrt{2.3 \alpha_{\rm turb} {\rm St}}$ 
\citep{OC07}.\footnote{ Here, the factor $\sqrt{2.3}$ assumes that the Stokes number 
of the smaller aggregate is 0.5 times that of the larger aggregate \citepalias{OMS+16}.  
For a collision of equal-sized aggregates, the prefactor is $\sqrt{3}$ \citep{OC07}. 
}
Since fragmentation dominates over sticking for $\Delta v \geq v_{\rm frag}$, 
aggregates can only grow to ${\rm St} \approx {\rm St}_{\rm frag}$, where 
\beqn
{\rm St}_{\rm frag} &=& \frac{ v_{\rm frag}^2}{2.3\alpha_{\rm turb}c_s^2}
\nonumber \\
&\approx& 3\times10^{-3} \pfrac{v_{\rm frag}}{0.5~\rm m~s^{-1}}^{2}\pfrac{\alpha_{\rm turb}}{3\times 10^{-4}}^{-1}
\pfrac{T}{30~\rm K}^{-1}.
\label{eq:Stfrag}
\eeqn
Using Equation~\eqref{eq:St}, the above expression can be translated into the 
maximum aggregate size in the fragmentation-limit growth,
\beqn
a_{\rm frag} 
&\approx& \frac{v_{\rm frag}^2 \Sigma_g}{4\alpha_{\rm turb} c_s^2\rho_{\rm int}}
\nonumber \\
&\approx& 300 \pfrac{v_{\rm frag}}{0.5~\rm m~s^{-1}}^{2} \pfrac{\alpha_{\rm turb}}{3 \times 10^{-4}}^{-1} \pfrac{\rho_{\rm int}}{0.6 ~\rm g~cm^{-3}}^{-1} 
\nonumber \\
&& \times  \pfrac{\Sigma_g}{10~\rm g~cm^{-2}}\pfrac{T}{30~\rm K}^{-1} ~\micron.
\label{eq:afrag}
\eeqn
Expressions similar to Equation~\eqref{eq:Stfrag} and \eqref{eq:afrag} can also be found 
in the literature \citep[e.g.,][]{BDB09,BKE12,PPSB17}. 
Equation~\eqref{eq:afrag} implies that 
the ten-fold decrease in $v_{\rm frag}$ due to the presence of 
CO$_2$ ice mantles leads to a 100-fold decrease in $a_{\rm frag}$ outside the CO$_2$ snow line.
In the first and second rows of Figure~\ref{fig:aSigma}, we compare $a$ and ${\rm St}$ 
from the simulations with ${a}_{\rm frag}$ and ${\rm St}_{\rm frag}$.   
For the models with CO$_2$ ice mantles, we find that ${a}_{\rm frag}$ well explains 
the aggregate size outside the CO$_2$ snow line.
This confirms our expectation that fragmentation induced by CO$_2$ mantle controls 
the aggregate size in the outer part of the disk.\footnote{ Strictly speaking, 
${\rm St}_{\rm frag}$ given by Equation~\eqref{eq:Stfrag} slightly overestimates 
the maximum Stokes number in the sintering zones.
In these regions, ${\rm St}$ is too small for 
the approximate expression $\Delta v \approx \sqrt{2.3 \alpha_{\rm turb} {\rm St}}$
for the turbulence-induced collision velocity to be valid. 
} 
Note that 
${\rm St}$ falls below ${\rm St}_{\rm frag}$ when dust growth is limited by radial drift. 
This can be seen in the models without CO$_2$ ice mantles.
}

{
Since $\alpha_{\rm frag} \propto \Sigma_{\rm g}$, 
the choice of $\gamma$ affects the radial slope of $a_*$
as can be seen in the top panels of Figure~\ref{fig:aSigma}.  
In the case of $\gamma = 1$, $a_*$ increases toward the disk center, and therefore
considerably deviates from $100~\micron$ at $r \la 50~\rm au$ even with CO$_2$ ice mantles.  
In contrast, the gas disk model with $\gamma = 0$ produces a flatter radial profile for $a_{\rm *}$
and thus ensures that CO$_2$-mantled grain aggregates have a size of 
$\sim 100~\micron$ all the way down to the CO$_2$ snow line.  
As we see in Section~\ref{sec:polari}, this difference affects 
the polarized emission from the inner $\sim 50~\rm au$ region of the disk. 
}

{ 
The third row of Figure~\ref{fig:aSigma} shows 
the radial profiles of the dust surface density $\Sigma_d$. 
In a quasi-steady flow of radially drifting aggregates, 
$\Sigma_d$ is inversely proportional to their drift speed, 
and hence to ${\rm St}$ when ${\rm St} < 1$ \citepalias[see Equation (25) of][]{OMS+16}. 
Thus, fragmentation induced by sintering causes enhancements of $\Sigma_g$ in the sintering zones, 
as already pointed out by \citetalias{OMS+16}.}
The models with CO$_2$ ice mantles predict
 a deep gap in $\Sigma_d$ between the H$_2$O and CO$_2$ snow lines. 
In these models, the inward mass flux of solids across the CO$_2$ snow line 
is low because the aggregates outside the snow line are small.
This causes a large deficit of solids between the H$_2$O and CO$_2$ snow lines, 
where the aggregates grow large and drift rapidly thanks to sticky H$_2$O mantles. 
This gap formation mechanism was also pointed out by \citet{PPSB17}.

The bottom panels of Figure~\ref{fig:aSigma} show the dust scale height $H_d$ as well as the gas scale height  $H_g  \approx 7(r/100~\rm au)^{1.2}~\rm au$.
The well-separated dust rings of the HL Tau disk suggest that the dust rings are vertically thin, with $H_d \approx 1~\au$ $(\sim H_g/7)$ 
at $r \approx 100~\au$ \citep{PDM+16}. 
With the choice of $\alpha_{\rm turb}  = 3 \times 10^{-4}$, all our models reproduce dust settling at the observed level.

\subsection{Synthetic Polarimetric Images}\label{sec:polari}
\begin{figure*}[t]
\centering
\includegraphics[width = 17.5cm, bb = 0 0 680 500]{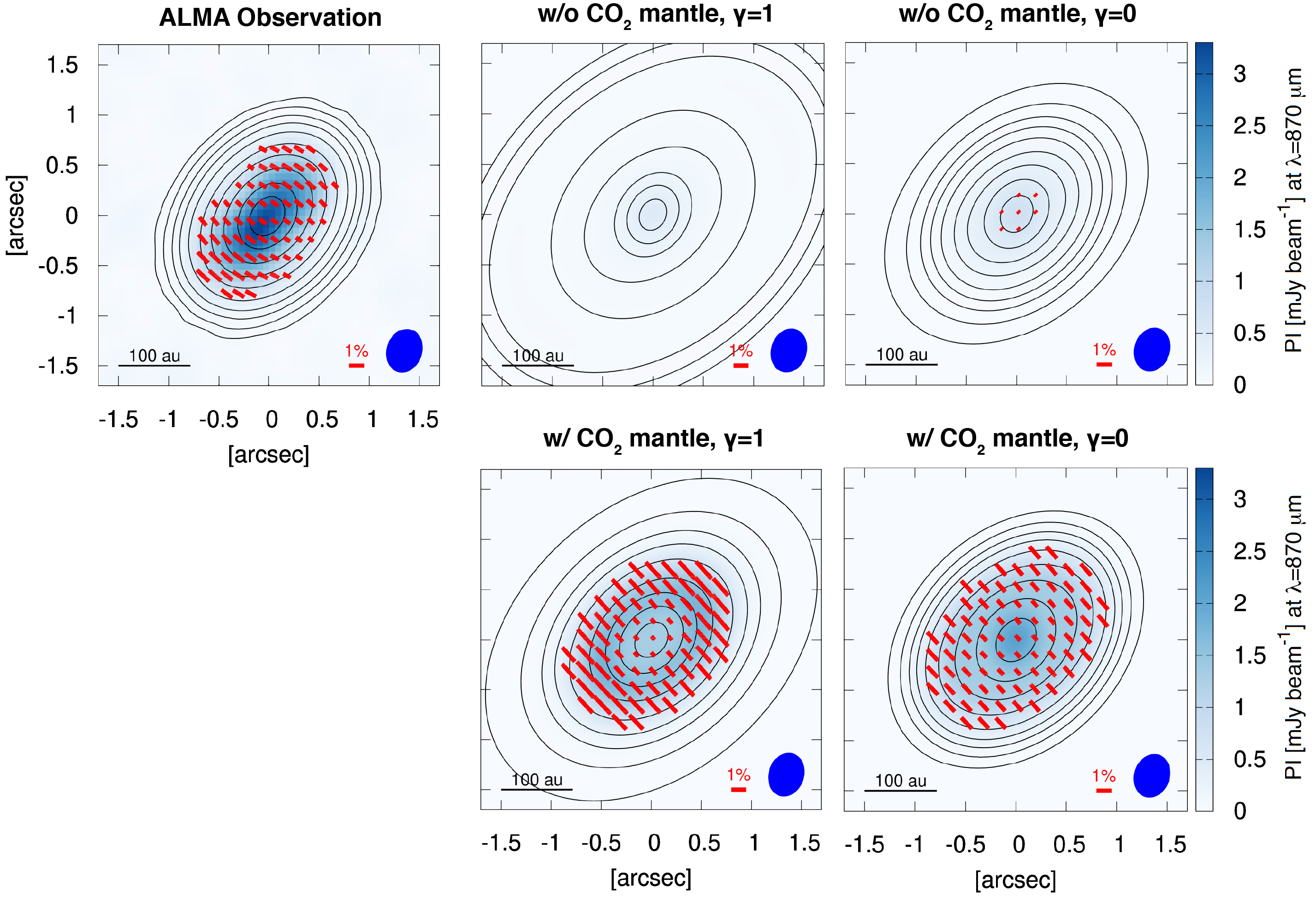}
\caption{{ Left column: ALMA polarimetric image of the HL Tau disk at $\lambda = 0.87~\rm mm$  \citep{SYL+17}. The color scale is the polarized intensity, and the red line segments show the direction and degree of polarization. }
Contours are the total intensity, with the contour intervals taken to be the same as in Figure 1 of \citet{SYL+17}.
{  
Center and right columns: synthetic polarimetric images at $\lambda = 0.87$ mm 
from four dust evolution models. 
The model images are after convolution with an elliptic Gaussian kernel of 
$0.\arcsec44\times 0.\arcsec35$, the beam size for the corresponding observation.
}
}
\label{fig:P}
\end{figure*}

\begin{figure*}[t]
\centering
\includegraphics[width = 17cm, bb = 0 0 800 550]{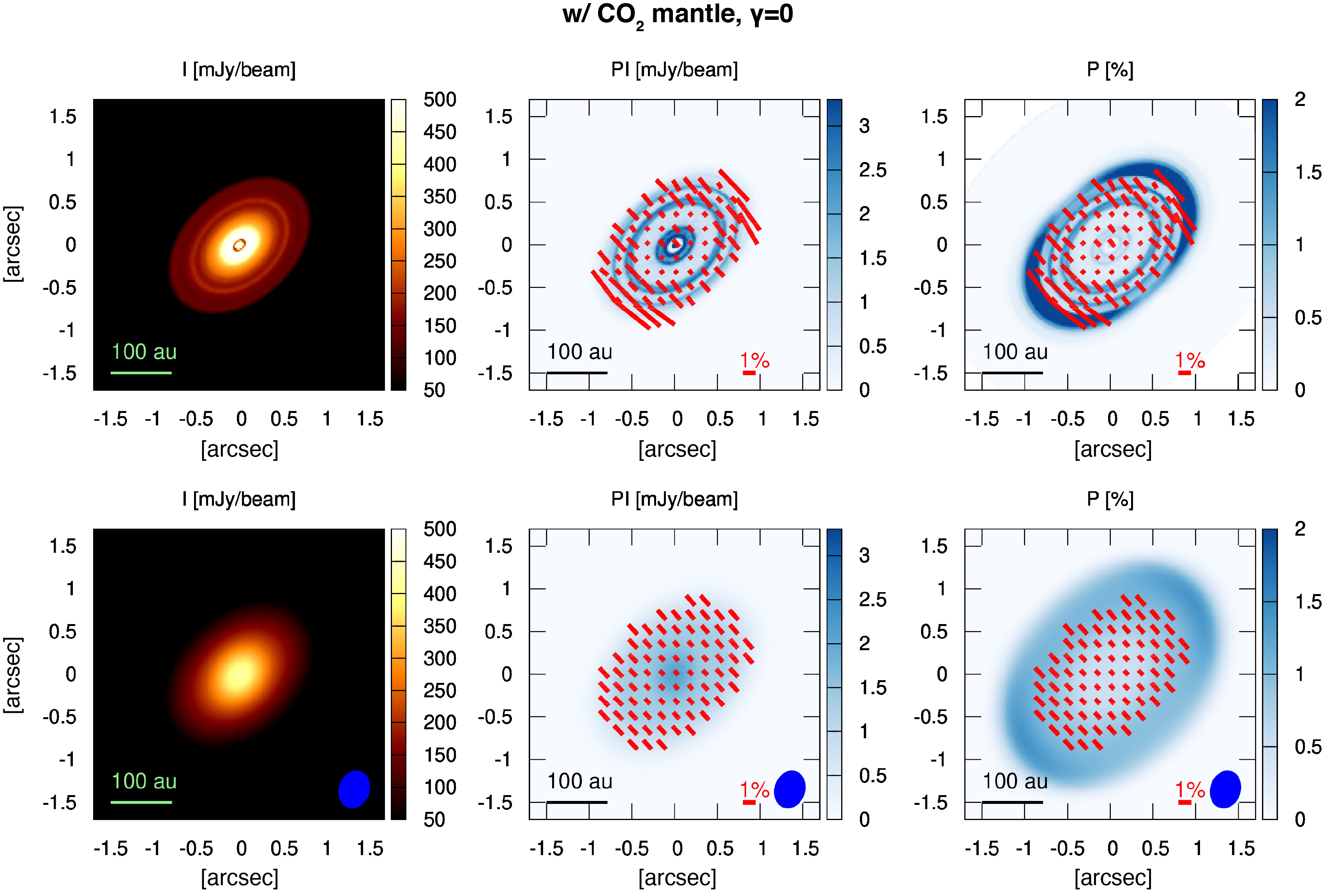}
\caption{ Model polarimetric image at $\lambda=$ 0.87 mm 
before and after Gaussian convolution (upper and lower panels, respectively)
from the model with CO$_2$ ice mantles and $\gamma = 0$. 
The beam shape is taken to be the same as that for the ALMA polarimetric image by \citet{SYL+17}.
The color scales for the left, center, and right panels show the total intensity $I$, polarized intensity $PI$, 
and the degree of polarization $P$, while the vectors in the center and right panels indicate the direction and 
degree of polarization. 
}
\label{fig:all1}
\end{figure*}

Figure~\ref{fig:P} presents the synthetic polarimetric images of the HL Tau disk
at wavelength $\lambda=$ 0.87 mm derived from the four models. 
{ The ALMA polarimetric image of the same object at the same wavelength
\citep{SYL+17} is also shown for comparison.
To allow direct comparison with the observation, we have produced the synthetic images
by convolved the ``raw'' images from radiation transfer calculations 
with an elliptic Gaussian kernel of size $0.\arcsec44\times 0.\arcsec35$, 
which is equal to the beam size for the corresponding ALMA observation (see Figure~\ref{fig:all1} for a raw image).}
At this wavelength, the observed polarimetric map shows 
a uniform polarization pattern with $PI \sim 1 ~\rm mJy~beam^{-1}$ 
and $P \sim 1\%$, where $PI$ and $P$ are the linear polarized intensity and 
the degree of linear polarization, respectively. 
{ We find that the models with CO$_2$ ice mantles successfully 
reproduce a similar unidirectional polarization pattern.
The mechanism responsible for this polarization is 
the self-scattering by $\sim 100~\micron$-sized aggregates. 
In contrast, the models without CO$_2$ ice mantles predict much weaker polarized emission.
This is not surprising since the aggregate size in those models is in much excess of $100~\micron$ over almost the entire part of the disk. }


{ Of the two models with CO$_2$ ice mantles, the one with $\gamma = 0$ 
better reproduces the observed polarized emission in the central region of $r \la 100~\rm au$.
In this region, the polarization degree of the observed emission is spatially uniform
to within a factor of $\approx 2$.
The model with $\gamma = 0$ successfully reproduces this feature 
because the aggregate size in the model stays at $\sim 100~\micron$ all the way down 
to the CO$_2$ snow line. 
In contrast, the model with $\gamma = 1$ predicts that
the aggregate size in the central part well exceeds $100~\micron$, 
and consequently underestimates the polarization degree there.
We note, however, the latter model could also provide an equall good match to the observation
if $v_{\rm frag}$, $\alpha_{\rm turb}$, and $\rho_{\rm int}$ are allowed to vary with $r$. 
For instance, according to Equation~\eqref{eq:afrag}, 
a model with $\Sigma_g \propto r^{-1}$ and $\alpha_{\rm turb} \propto r^{-1}$ 
would yield the same radial dependence for $a_{\rm frag}$ as the model with radially constant 
$\Sigma_g$ and $\alpha_{\rm turb}$. 
}

{ Given their relatively low angular resolutions ($\approx 60~\rm au$), 
the previous polarimetric observations of the HL Tau disk were not able to 
resolve any substructure on 10 au scales, where multiple dust rings are visible \citep{ALMA+15}. 
However, for future high-resolution polarimetric observations, 
it is important to predict how the polarization pattern of the disk would look like
on the substructure scale. 
We show in the first row of Figure~\ref{fig:all1}
 the raw maps of $I$, $PI$, and $P$ for the model with CO$_2$ ice mantles and $\gamma = 0$
 directly obtained form the radiative transfer calculation.   
For comparison, we also show in the first row of Figure~\ref{fig:all1}
the maps after Gaussian smoothing with resolution $0.\arcsec44\times0.\arcsec35$, 
which are  essentially the same as the lower right panel of Figure~\ref{fig:aSigma}.
The bright rings visible in the total intensity map before smoothing (the upper left panel of Figure~\ref{fig:all1}) correspond to the pileups of dust in the sintering zones.  
The map for $PI$ (the upper center panel of Figure~\ref{fig:all1}) indicates that 
these bright rings are also responsible for the dominant fraction of the polarized emission.
The emission from the dust rings has high $PI$ and high $P$,
because the aggregate size in the sintering zones is particularly close to $100~\micron$.
However, this correlation between $P$ and $I$ is  model-dependent; 
if $a_*$ in the model were smaller by a factor of 3 at all $r$, 
the degree of polarization would be lower in brighter regions than in darker regions. 
In any case, future polarimetric observations of the HL Tau disk with higher angular resolutions 
may test these predictions.
More detailed comparisons between our models and the long-baseline observations of HL Tau
are presented in Section~\ref{sec:ring}.
}

We also performed radiative transfer calculations at longer wavelengths 
and confirmed that polarization due to dust self-scattering diminishes
at $\lambda \gg 1~\rm mm$, even with CO$_2$ ice mantles.
This is in qualitative agreement with the fact 
that the uni-directional polarized emission is not observed at $\lambda  = 3.1~\rm  mm$  \citep{KTP+17}.
{ We avoid more quantitative comparisons at these wavelengths
because our current models do not include  mechanisms that can produce 
an azimuthal polarization pattern as seen in the observed images. 
}

\section{Discussion}\label{sec:discussion}
\subsection{Implications for the Ring--Gap Substructure and Millimeter 
Spectral Slope of the HL Tau Disk}\label{sec:ring}
\begin{figure*}[t]
\centering
\resizebox{\hsize}{!}{\includegraphics{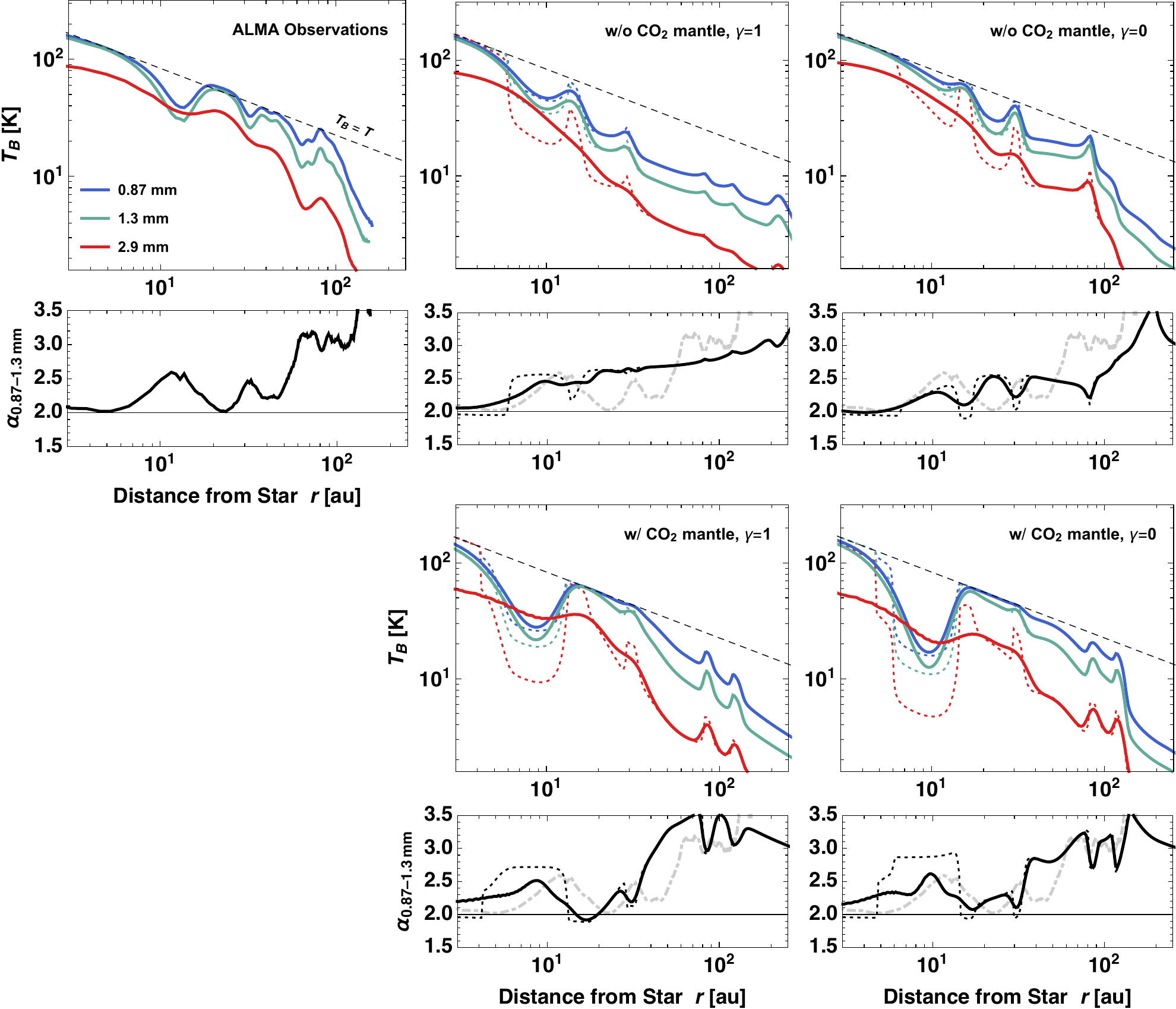}}
\caption{ Left column: azimuthally averaged millimeter emission profiles 
from the long-baseline observations of the HL Tau disk 
(\citealt{ALMA+15}; see \citetalias{OMS+16} for azimuthal averaging).  
The upper panel shows the Planck brightness temperatures $T_B$ at $\lambda =$ 0.87, 1.3, and 2.9 mm,
and the lower panel shows the spectral index at $\lambda =$ 0.87--1.3 mm, $\alpha_{0.87\textrm{--}1.3~\rm mm}$.
The dashed line shows the power-law disk temperature profile assumed in this study. 
Center and right columns: profiles of $T_B$ and $\alpha_{0.87\textrm{--}1.3~\rm mm}$ from four models.
The dotted lines show the raw profiles directly obtained from simulation results,
while the solid lines show the profiles obtained by Gaussian smoothing the model intensities 
at the spatial resolutions of the observations. The gray dot-dashed line shows 
the spectral index profile from the observations for comparison. 
}
\label{fig:TB}
\end{figure*}
So far we have focused on the polarized emission of the HL Tau disk on a $100 ~\rm au$ scale.
It is also interesting to see what our model implies for the ring--gap intensity profiles
observed in the earlier ALMA observations \citep{ALMA+15}.  
{ In the center and right panels of Figure~\ref{fig:TB}, 
we plot the radial profiles of the Planck brightness temperatures $T_B$ 
at $\lambda = $ 0.87, 1.3, and 2.9 mm (corresponding to ALMA Bands 7, 6, and 3)
obtained from our dust evolution models. 
Here, the three-band brightness temperatures are calculated 
using Equations~(21) and (22) of \citetalias{OMS+16}, 
which assume slab geometry and neglect scattering. 
This approach gives an accurate estimate for the total intensities 
unless the disk is optically thick and has a high albedo \citep{RL79,BDZ18,
L190400333,ZZJ1904.02127}. 
This is the case at least  in the model with CO$_2$ mantles and $\gamma = 0$, 
where the aggregates in the optically thick sintering zones have a relatively low albedo of $\sim 0.4$.
The solid and dotted lines in Figure~\ref{fig:TB} show the profiles before and after Gaussian smoothing
at the spatial resolutions of the corresponding ALMA observations.
The local bumps visible in the model $T_B$ profiles correspond 
to the dust rings in the sintering zones.
For comparison, we also show in the left panel Figure~\ref{fig:TB} 
the radial brightness temperature profiles of the HL Tau disk
derived from the ALMA high-resolution images \citep{ALMA+15}.
 \citetalias{OMS+16} produces these profiles by azimuthally averaging the observed maps
(see Section 2.1 of \citetalias{OMS+16} for more details).}

The most prominent feature in the intensity profiles 
caused by CO$_2$ ice mantles is a deep gap at $r \approx $ 10 au.
Interestingly, the location of this deep gap coincides with that of HL Tau's 
innermost gap \citep{ALMA+15}. 
In these models, the deep gap is caused by the rapid growth and depletion of 
H$_2$O-mantled grain aggregates between the H$_2$O and CO$_2$ 
snow lines as described in Section 3.1.  
{ 
The models with CO$_2$ ice mantles generally predict that
the 10 au gap is the deepest among the multiple dust rings,
in qualitative agreement with the ALMA images of the HL Tau disk. 
However, these models overestimate the depth of the 10 au gap, 
implying that further tuning of model parameters is needed to achieve a more quantitative match. 
Sintering also produces a 10 au gap \citetalias{OMS+16} as seen in the models without CO$_2$ ice mantles. However, this sintering-induced  is substantially shallower than the 10 au gap in the models with CO$_2$ ice mantles. }

Our results also offer an important interpretation for
the millimeter spectral slope of the HL Tau disk. 
We define the spectral index at $\lambda = $ 0.87--1.3 mm (i.e., between ALMA Bands 7 and 6)
as $\alpha_{0.87\textrm{--}1.3~\rm mm} \equiv \ln(I_6/I_7)/d\ln(\nu_6/\nu_7)$, 
where $I_{6,7}$ and $\nu_{6,7}$ are the intensities and frequencies at $\lambda = $ 1.3 and 0.87 mm.
{ The radial profiles of $\alpha_{0.87\textrm{--}1.3~\rm mm}$ from the observations and models
are shown in Figure~\ref{fig:TB}  
(for comparison, the observed profile is also overplotted in the panels for the model profiles).}
The profile from the observations shows $\alpha_{0.87\textrm{--}1.3~\rm mm} \sim 2$--2.5 
at $r \la 50~\au$ and $\alpha_{0.87\textrm{--}1.3~\rm mm} \sim 3$ farther out 
\citep[see also][]{ZBB15}.
A millimeter spectral index of $\sim 2$--3 is commonly attributed { either to 
dust growth to millimeter sizes, or to a moderately large optically thickness} \citep[e.g.,][]{RTN+10}.
{ The model with CO$_2$ ice mantles and $\gamma = 0$} 
suggests that the latter interpretation applies to the HL Tau disk, except at the 10 au gap.  
In fact, this model predicts $a_* \sim 100~\micron$ at $r \ga 50~\au$ but still
successfully reproduces $\alpha_{0.87\textrm{--}1.3~\rm mm}$ in this outer part 
(optically thin emission from $100~\micron$ particles 
would have $\alpha_{0.87\textrm{--}1.3~\rm mm} \sim 3.5$).
Only the spectral index of the optically thin emission 
from the 10 au gap reflects the presence of mm-sized aggregates 
between the H$_2$O and CO$_2$ snow lines.
{ 
Our interpretation is consistent with the optical depth estimates by \citet{CHC+16},
who suggest the HL Tau disk is marginally optically thick (with optical depth $\sim 0.3$--3) 
at $\lambda \la 0.87$ and 1.3 mm (see their Figure 3). 
}

{ 
\subsection{Parameter Dependence}\label{sec:dependence}
We have demonstrated that the low stickiness of CO$_2$-mantled 
grain aggregates reasonably explains the abundant presence of $100~\micron$-sized particles in disks.
However, it is clear from Equation~\eqref{eq:afrag} that the maximum particle size 
depends not only on the sticking threshold $v_{\rm frag}$ but also 
on other parameters such as turbulence strength and aggregate porosity. 
In this subsection, we show that a low fragmentation threshold is still the most likely 
explanation for $100~\micron$-sized particles
in the particular case of the HL Tau disk.

A key constraint comes from the evidence for dust settling in the HL Tau disk. 
The distinct morphology of the observed dust rings suggests 
$H_d \la 1~\rm au$ at $r \approx 100~\rm au$  in this disk \citep{PDM+16}, 
which translates into $H_d \la H_g/7$ in our disk model (see also Section~\ref{sec:radial}).
This can be used to constrain turbulence strength $\alpha_{\rm turb}$. 
From Equation~\eqref{eq:Hd} with $\alpha_{Dz} \approx 0.3 \alpha_{\rm turb}$, 
one has $H_d/H_g \approx (0.3\alpha_{\rm turb}/{\rm St})^{1/2}$ for $H_d \ll H_d$.  
When dust growth is limited by turbulence-induced fragmentation, 
${\rm St}$  is given by Equation~\eqref{eq:Stfrag}, and we obtain 
\beqn
\alpha_{\rm turb} &\approx& \frac{H_d}{H_g}\frac{v_{\rm frag}}{c_s}
\nonumber \\
&\approx& 2\times 10^{-4}\pfrac{7H_d}{H_g}
\pfrac{v_{\rm frag}}{0.5~\rm m~s^{-1}}\pfrac{T}{30~\rm K}^{-1/2}.
\label{eq:alpha_limit}
\eeqn
Equation~\eqref{eq:alpha_limit} can in turn be used to eliminate $\alpha_{\rm turb}$
from the maximum aggregate size $a_{\rm frag}$ given by Equation~\eqref{eq:afrag}. 
The result is
\beqn
a_{\rm frag} 
&\approx& \pfrac{H_d}{H_g}^{-1}\frac{v_{\rm frag} \Sigma_g}{4 c_s \rho_{\rm int}} 
\nonumber \\
&\approx& 400\pfrac{7H_d}{H_g}^{-1} \pfrac{v_{\rm frag}}{0.5~\rm m~s^{-1}}
\pfrac{\rho_{\rm int}}{0.6 ~\rm g~cm^{-3}}^{-1}
\nonumber \\
&& \times  \pfrac{\Sigma_g}{10~\rm g~cm^{-2}}\pfrac{T}{30~\rm K}^{-1/2} ~\micron.
\label{eq:afrag_limit}
\eeqn
The weak temperature dependence 
of the right-hand side of Equation~\eqref{eq:afrag_limit} is negligible 
as long as we consider $r \sim $10--100 au. 

Equation~\eqref{eq:afrag_limit} indicates that
any model reproducing the two observational constraints 
$H_d \la H_g/7$ and $a_{\rm frag} \sim 100~\micron$
must satisfy the condition  
\beq
\pfrac{v_{\rm frag}}{1~\rm m~s^{-1}} \pfrac{\rho_{\rm int}}{1~{\rm g~cm^{-3}}}^{-1}
\pfrac{\Sigma_g}{10~{\rm g~cm^{-2}}} \la 0.1.
\label{eq:vfrag_cond}
\eeq
Any realistic dust aggregate has $\rho_{\rm int} \la 1$--3$~\rm g~cm^{-3}$, 
and the high millimeter flux from the HL Tau disk points to
$\Sigma_{\rm g} \sim 10~\rm g~cm^{-2}$ at $r \sim 50$--100 au
unless the dust-to-gas mass ratio is well above 0.01 
(e.g., \citealt{KLM11,KLMW15}; \citealt{PDM+16}; see also \citetalias{OMS+16}). 
Therefore, we conclude that only fragmentation thresholds as low as $\la 0.1$--$1~\rm m~s^{-1}$
can explain $H_d \la H_g/7$ and $a_{\rm frag} \sim 100~\micron$ simultaneously.

The requirement $v_{\rm frag} \la 0.1$--$1~\rm m~s^{-1}$
translates into monomers sizes of $a_{\rm mon} \ga 1~\micron$ for CO$_2$-mantled grain aggregates 
and $a_{\rm mon} \ga 10~\rm \micron$ for H$_2$O ice aggregates. 
The former is more consistent with the maximum grain size
$\sim 1~\micron$ for the HL Tau envelope
inferred from near-infrared polarimetry \citep{LFT+04,MOPI08},
suggesting that  CO$_2$-mantles offer a more reasonable explanation for the origin of 
100~$\micron$-sized aggregates.

\subsection{Are Aggregates Compact or Fluffy?}\label{sec:porous}
We have assumed that the aggregates in the HL Tau disk have a moderately high filling factor of 50\%.
If the aggregates are fluffy, i.e., $\rho_{\rm int} \ll 1~\rm g~cm^{-3}$,
Equation~\eqref{eq:vfrag_cond} demands an even lower value of $v_{\rm frag}$. 
The question is then whether such a low fragmentation threshold is realistic for fluffy aggregates.
Although aggregate collision simulations do show a trend of decreasing $v_{\rm frag}$ with increasing aggregate porosity \citep{WTS+09,GRU16}, it is unclear at the present whether this trend can compensate for 
the decrease of $\rho_{\rm int}$ in Equation~\eqref{eq:vfrag_cond}. 

A more fundamental question about fluffy aggregates is whether they can produce strongly polarized scattered light.
The effective medium approach adopted in this study is inapplicable to fluffy aggregates, in particular to fractal aggregates of fractal dimension $\la 2$ \citep{TT18}.
For such aggregates, a more accurate scattering theory, such as the modified mean field theory \citep{TTO+16,TTM+19,TT18}, should be used to obtain optical properties.
We leave this extension for future publication.\footnote{
{ Our preliminary calculations using the modified mean field theory seem to 
show that {polarized intensity due to self-scattering}
diminishes as the aggregate porosity increases (Tazaki et al., in prep.).}}

From mechanical point of view, it is also unclear how icy aggregates in the disk could become compact.
\citet{OTKW12} and \citet{KTOW13b} showed that neither aerodynamical nor collisional compression 
is efficient enough to compress icy aggregates in protoplanetary disks to a filling factor of $\ga 0.1$. 
However, these previous studies assumed that the ice aggregates are sticky and grow without fragmentation. It is yet to be explored how the porosity of aggregates evolves 
when they experience highly destructive collisions and subsequently reaccrete a large amount of small fragments. 
Poorly sticky aggregates may also lose a high porosity through bouncing collisions \citep{WGBB09,ZODH11}. 
}

\section{Conclusions}\label{sec:summary}
{ We have demonstrated that aggregates with a low sticking efficiency  
can reasonably produce the uniformly polarized emission seen in the HL Tau disk. 
In our scenario, the growth of the icy aggregates is primarily limited by 
the mantles of nonsticky CO$_2$ ice, and to a less extent by aggregate sintering.  }
Dust aggregates still can grow beyond 1 mm between the H$_2$O and CO$_2$ snow lines, 
where CO$_2$ ice mantles are absent and the sticky H$_2$O mantle facilitates dust growth.
The efficient growth and subsequent rapid infall of the dust in this region might explain the 10 au dust gap in the HL Tau disk.
The models with CO$_2$ ice mantles also suggest that the low spectral index of the HL Tau disk at 0.87--1.3 mm 
primarily reflects the optical thickness at this wavelength, not the size of opacity-dominating particles, except in the 10 au gap.

If our interpretation is correct, the polarimetric observation of HL Tau may provide the first evidence
for the dust fragmentation barrier induced by CO$_2$ ice \citep{MTJW16a,MTJW16b,PPSB17}. 
The CO$_2$-induced fragmentation barrier may also solve the long-standing problem of 
dust retention in protoplanetary disks over several million years, 
which requires substantial particle fragmentation to slow down their radial infall \citep{DD05,BDB09}.
Future applications of our model to other disks with a similar (sub)millimeter polarization pattern
will enable us to better understand the role of CO$_2$ ice on dust evolution in protoplanetary disks.

{ 
Although we have assumed that H$_2$O ice is stickier than CO$_2$ ice, 
recent laboratory experiments suggest that H$_2$O ice might also be poorly sticky 
at low temperatures \citep{GSK+18,MW19}. 
The two scenarios would equally well reproduce the uniformly polarized submillimeter emission
from many protoplanetary disk, but the latter scenario would predict no deep dust gap 
between the H$_2$O and CO$_2$ snow lines. 
Therefore, the two scenarios could be tested by observing the dust continuum emission from 
the vicinity of these snow lines in different protoplanetary disks.

Bouncing collisions, which we neglected in this study, could also cause 
a low sticking thresholds for icy aggregates and should be taken into account in future work.
Compaction of aggregates through bouncing collisions \citep{WGBB09,ZODH11} could also justify
our assumption that icy aggregates have a relatively high filling factor.

It should be noted that whether CO$_2$ ice suppresses dust growth depends on 
how CO$_2$ ice is distributed inside grains. As already mentioned in Section~\ref{sec:grain},
the sticking efficiency of mixtures CO$_2$ and H$_2$O ices would be similar to that of pure H$_2$O ice if CO$_2$/H$_2$O $\sim 0.1$ \citep{MTJW16b}.
Therefore, it is possible that CO$_2$ ice does not produce $\sim 100~\micron$ grains in every protoplanetary disk.
Such a  variety can also be expected from recent ALMA observations of disk substructures. While not all observed disks possess substructures clearly associated with snow lines \citep{LPH18,HAD18,VDD19}, some systems (HD 135544 B, HD 169142, and HD 97048) do appear to have a deficit of dust interior to the CO$_2$ snow line (see Figure 9 of \citealt{VDD19}). 

We also note here that sintering may not lead to the formation of dust rings in all protoplanetary disks. \citet{S11b} suggests that icy aggregates can be sintered over a wide region of a protoplanetary disk if they are temporally transported to the disk surface and get heated there. If this is the case, sintering would not lead to local concentration of dust in the radial direction. This might be another reason why not all spatially resolved protoplanetary disks seem to have dust rings associated with snow lines. The HL Tau disk is perhaps an ideal environment for sintering-induced ring formation because vertical mixing of dust is ineffective in this disk \citep{PDM+16}. 
}

Finally, we note that the models presented in this paper do not address 
the origin of the polarized emission of the HL Tau disk at 3.1 mm.
At this wavelength, the polarimetric image exhibit no unidirectional polarization pattern,
only showing azimuthal polarization \citep{KTP+17}. 
Mechanisms that can produce azimuthally polarized emission,
such as radiative { and aerodynamic grain alignment \citep{TLN17,KOT19}}, 
must also be taken into account to fully understand the polarized emission from the HL Tau disk, and perhaps from other protoplanetary disks as well.   

\acknowledgments
{ We are grateful to Ian Stephens and Akimasa Kataoka for kindly providing us with the FITS image of the HL Tau disk \citep{SYL+17}.
We also thank Kota Higuchi, Kenji Furuya, Takahiro Ueda, Carlos Carrasco-Gonz\'{a}lez,
Zhaohuan Zhu, Munetake Momose, and Tomoko Suzuki for useful discussions, and 
the anonymous referee for comments that significantly improved the paper.} 
S.O.~was supported by JSPS KAKENHI Grant Numbers JP16H04081, JP16K17661, JP17K18812, JP18H05438, and JP19K03926. R.T.~was supported by a Research Fellowship for Young Scientists from the Japan Society for the Promotion of Science (JSPS) (17J02411).

\bibliography{myrefs_190312.bib}

\end{document}